\documentclass[draft]{agujournal2019}
\usepackage{url} 
\usepackage{lineno}
\usepackage{amsmath, amsfonts}
\usepackage{microtype}
\usepackage[inline]{trackchanges} 
\usepackage{soul}
\newcommand{\rr}{\mathbb{R}}
\newcommand{\var}{\textsc{var}}
\newcommand{\E}{\textsc{E}}

\newcommand{\vx}{{\mathbf x}}

\newcommand{\vv}{{\mathbf v}}

\newcommand{\va}{{\mathbf a}}
\newcommand{\vu}{{\mathbf u}}

\newif\ifnotesw \noteswtrue

\usepackage{setspace}
\usepackage{etoolbox}
\BeforeBeginEnvironment{equation}{ \begin{singlespace}\vspace*{-\baselineskip}\vspace*{-0.2cm}}
\AfterEndEnvironment{equation}{\end{singlespace} \vspace*{-0.3cm}\noindent\ignorespaces}

\BeforeBeginEnvironment{equation*}{\begin{singlespace}\vspace*{-\baselineskip} \vspace*{-0.2cm}}
\AfterEndEnvironment{equation*}{\end{singlespace}\vspace*{-0.3cm}\noindent\ignorespaces}

\BeforeBeginEnvironment{align}{\begin{singlespace}\vspace*{-\baselineskip} \vspace*{-0.2cm}}
\AfterEndEnvironment{align}{\end{singlespace}\vspace*{-0.3cm}\noindent\ignorespaces}

\BeforeBeginEnvironment{align*}{\begin{singlespace}\vspace*{-\baselineskip}\vspace*{-0.2cm}}
\AfterEndEnvironment{align*}{\end{singlespace}\vspace*{-0.3cm}\noindent\ignorespaces}

\AtBeginEnvironment{tabular}{\singlespacing}

\usepackage{titlesec}

\titlespacing\section{0pt}{0pt}{0pt plus 2pt minus 2pt}
\titlespacing\subsection{0pt}{0pt}{0pt plus 2pt minus 2pt}
\titlespacing\subsubsection{0pt}{0pt}{0pt plus 2pt minus 2pt}

\draftfalse

\journalname{JGR: Solid Earth}

\begin{document}

%
%


\title{Sensitivity Analysis in the Presence of Intrinsic Stochasticity for Discrete Fracture Network Simulations}

%
%




\authors{A.C. Murph\affil{1}, J.D. Strait\affil{1}, K.R. Moran\affil{1}, J.D. Hyman\affil{2,3}, H.S. Viswanathan\affil{2}, and P.H. Stauffer\affil{2}}


\affiliation{1}{Statistical Sciences (CCS-6), Computer, Computational, and Statistical Sciences
Division, Los Alamos National Laboratory, Los Alamos, NM, 87545}
\affiliation{2}{Energy and Earth System Science (EES-16), Earth and Environmental Sciences Division, Los Alamos National Laboratory, Los Alamos, NM, 87545}
\affiliation{3}{Department of Geology and Geological Engineering, Colorado School of Mines, Golden, CO 80401}





\correspondingauthor{A.C. Murph}{murph@lanl.gov}



\begin{keypoints}
\item DFN simulators where the aleatoric uncertainty changes for different inputs makes several standard statistical methods inadmissible.
\item We handle each type of uncertainty separately and strategically collect simulator runs at input values that minimize the predictive error. 
\item The major drivers of variance in the outputs are the input parameters that govern the overall transmissivity of the system.
\end{keypoints}

%
%

%
%


\begin{abstract}
Large-scale discrete fracture network (DFN) simulators are standard fare for studies involving the sub-surface transport of particles since direct observation of real world underground fracture networks is generally infeasible.  While these simulators have seen numerous successes over several engineering applications, estimations on quantities of interest (QoI) — such as breakthrough time of particles reaching the edge of the system — suffer from two distinct types of uncertainty.  A run of a DFN simulator requires several parameter values to be set that dictate the placement and size of fractures, the density of fractures, and the overall permeability of the system; uncertainty on the proper parameter choices will lead to some amount of uncertainty in the QoI, called epistemic uncertainty.  Furthermore, since DFN simulators rely on stochastic processes to place fractures and govern flow, understanding how this randomness affects the QoI requires several runs of the simulator at distinct random seeds.  The uncertainty in the QoI attributed to different realizations (i.e. different seeds) of the same random process (i.e. identical input parameters) leads to a second type of uncertainty, called aleatoric uncertainty.  In this paper, we perform a Sensitivity Analysis, which directly attributes the uncertainty observed in the QoI to the epistemic uncertainty from each input parameter and to the aleatoric uncertainty.  Beyond the specific takeaways on which input variables influence uncertainty in the QoI the most, a major contribution of this paper is the introduction of a statistically rigorous workflow for characterizing the uncertainty in DFN flow simulations that exhibit heteroskedasticity. 
\end{abstract}

\section*{Plain Language Summary}
Observed heteroskedasticity in Discrete Fracture Network (DFN) simulators, where the ``random white noise" changes for different inputs, makes several standard statistical methods inadmissible.  We fit a statistical model on DFN simulation data that directly addresses this issue with the aim of performing a Sensitivity Analysis (SA).  In a SA, uncertainty on some Quantity of Interest (QoI) -- which for this application is the log-time until particles simulated through a discrete fracture network reach the surface -- is attributed to uncertainty on the DFN simulation input parameters.  This analysis leads to a better understanding on which simulation input parameters lead to the greatest variation in the QoI, and how much of the variation on this QoI is noise that cannot be attributed to any input parameters (i.e. white noise).

\section{Introduction}
The flow of fluids (liquids and gases) and associated transport of chemical species through low perm rocks in the subsurface primarily occurs within interconnected networks of fractures \cite{viswanathan2022from}.  Modeling transport thorough fracture networks is the focus of numerous applications including aquifer management, hydrocarbon extraction, detection of nuclear weapons testing, and the long-term storage of spent nuclear fuel ~\cite{follin2014methodology,hyman2016understanding,jenkins2015state,kueper1991behavior,middleton2017shale,national1996rock,neuman2005trends,selroos2002comparison,vanderkwaak1996dissolution,jordan2015, carrigan2022}. 
Since direct observation of fracture networks in subsurface rock is generally infeasible, computational models of flow are constructed stochastically~\cite{neuman2005trends,zhang2006stochastic}.
Beyond the uncertainty inherent to performing predictions with scant real-world data, each computational methodology has its own parameters which need to be populated and come with their own uncertainty. 
One methodology in use today are discrete fracture networks (DFNs) models where fractures and the networks that they form are explicitly represented~\cite{berrone2013pde,berrone2015parallel,cacas1990modeling,davy2013model,davy2010likely,de2004influence,dershowitz1999derivation,dreuzy2012influence,erhel2009flow,hyman2014conforming,hyman2015dfnworks,pichot2012generalized,mustapha2007new,ahmed2015control,ahmed2015three,antonietti2016mimetic,berrone2018advanced,Flemisch2016,fum,ODS,lagrange,manzoor2018interior,Schwenck2015,hyman2022flow}.
DFN simulations require users to set values for several detailed parameters for fracture size, orientation, and permeability of the fractures, all of which may greatly affect the outcome of the simulation.  
While such high fidelity DFN simulations have seen repeated success in studying subsurface flow, several studies have acknowledged the numerous layers of uncertainty involved in simulating flow through fractured media using the DFN methodology \cite{bonnet2001scaling, berkowitz2002, national2021characterization, hyman2019characterizing, strait2023,osthus2020probabilistic, omalley2018efficient,viswanathan2022from}.
Furthermore, because each network is generated stochastically rather than deterministically, identical inputs may lead to several different outcomes.  
 In summation, it difficult to parse whether variation in an outcome is due to a varying parameter input or due to intrinsic stochasticity.

There are multiple input parameters required to perform a sub-surface gas transport simulation using a DFN.  Fitting these input parameters demands a careful analysis of the process one is trying to emulate, and a thorough validation of the model post-fit.  Validating a statistical model or large-scale simulation is not about ``proving it to be true,'' but rather about corroborating its uses and thoroughly understanding its limitations.  This latter goal is generally undertaken using an Uncertainty Analysis (UQ), where the uncertainty in a model's output is quantified. This can (and often should) be taken a step further with a Sensitivity Analysis (SA), where the uncertainty quantified in a model's output is then apportioned to uncertainties in the model's input \cite{wiley2007}.  
In large-scale simulations, the sheer complexity of the modeled system often requires several tuning parameters that may be difficult to rigorously estimate, which makes an SA a paramount aim.  Indeed, several modern applications involving large-scale simulations have attempted to do just this, such as in nuclear waste storage safety studies \cite{iooss2006}, the study of neutron spectrums \cite{jacques2006}, pollutant transport modeling in aquifers \cite{volkova2008}.  A standard for these sorts of analyses are Sobol Indices \cite{sobol2001}, which assess the overall model uncertainty that can be attributed to each input (a global SA).  This is in contrast to examining how uncertainty can be attributed at specific areas in the input space (a local SA).  In this paper, any use of the abbreviation SA will refer to a global SA using Sobol indices.  For an overview of other SA methods, see \citeA{wiley2007}.




This paper proposes a rigorous workflow to characterize the uncertainty of an ensemble of three-dimensional discrete fracture network simulations.  For our simulations, we resolve flow and gas transport through semi-generic fracture networks using the {\sc dfnWorks} computational suite.  We separate the epistemic uncertainty attributed to changes in the simulation parameters from the aleatoric uncertainty that is treated as variation caused for different random seeds set across the ensemble.  Initial observations on DFN simulations have shown that the noise attributed to aleatoric uncertainty is heteroskedastic, meaning the amount of noise caused by a varying random seed changes for different input parameters.  As an example of heteroskedasticity, and a potential explanation for why it is observed in DFN simulations, consider the variable that governs the density of fractures: $P_{32}$.  For large $P_{32}$ values, there may be a higher probability that several paths direct paths exist from the source of particle flow to the edge of the system.  This would lead to overall less variation in breakthrough time for particles for different random seeds, since most of the ensemble would simulate the direct, uncongested transport of all particles out of the system.  In contrast, for low $P_{32}$ values there may be a higher probability that only one path exists from source to exit, making the breakthrough time abnormally high for some random seeds.  This would lead to very high variability across the ensemble for low $P_{32}$.  While this example is currently speculative for the application in this paper, it serves as an example of how the aleatoric uncertainty attributed to changes in a random seed may increase or decrease for different input parameters.

Due to runtime considerations for high-fidelity DFN simulators, we build a heteroskedastic Gaussian Process (GP) emulator on a limited number of simulation runs and perform a formal Sensitivity Analysis (SA) using this model.  
A heteroskedastic GP emulates a process with non-constant noise by way of two joint emulators, as described in Section \ref{subsec:het_joint_gauss_process}.  
To ensure that this emulator accurately represents the complicated data generating process, these limited simulation runs are first taken at input locations using a Latin hypercube design so as to fill the space, then augmented with additional runs up to the computational budget according to a sequential design.  As discussed in later sections, this sequential design selects new design points (input parameter ``locations" at which to perform simulations) in a way that minimizes average predictive error of the GP fit to the current training sample's observed breakthrough times; this sampling scheme allows for both replications at existing design points and for entirely new input parameter locations.  This experimental design yields a well-fit emulator, which is then used to quantify both types of variance in our output: the variance due to the input parameters and the variance attributed to changes in the random seed.

High-level findings are that the major drivers of variance in the outputs are the input parameters that govern the overall transmissivity of the system.  Perhaps unexpectedly, there is not sufficient evidence that the density of fractures interacts with the aleatoric uncertainty, as hypothesized earlier.  In fact, although we certainly have evidence that this random noise is heteroskedastic, we do not find sufficient evidence that this heteroskedasticity is correlated with changes in the input space.  The most likely explanation is that these changes themselves are random. 

The remainder of the paper is organized as follows.  In Section \ref{sec:simulations}, we describe our ensemble of DFN simulations and quantities of interest.  In Section \ref{sec:methodology}, we provide an overview of the methods used to fit the heteroskedastic GP, chose design points, and analyze the variance.  In Section \ref{sec:experiment}, we perform our analysis.

\section{Simulator Inputs and Outputs}\label{sec:simulations}

In this section, we describe our flow and transport simulations in three-dimensional fracture networks. 

\subsection{Discrete Fracture Network Simulations}

We use the high-fidelity three-dimensional discrete fracture network modeling suite {\sc dfnWorks}~\cite{hyman2015dfnworks} to generate each DFN, solve the steady-state flow equations, and simulate transport using particle tracking. 
{\sc dfnWorks} combines the feature rejection algorithm for meshing ({\sc fram})~\cite{hyman2014conforming}, the LaGriT meshing toolbox~\cite{lagrit2011}, the parallelized subsurface flow and reactive transport code {\sc pflotran}~\cite{lichtner2015pflotran}, and {\sc dfnTrans} particle tracking method ~\cite{makedonska2015particle,painter2012pathline,hyman2019matrix}.  
{\sc fram} is used to generate three-dimensional fracture networks.  
LaGriT is used to create a computational mesh representation of the DFN in parallel using the near-Maximal Poisson Disc sampling method (nMAPS)~\cite{krotz2022maximal}.
The massively parallel flow and reactive transport code  {\sc pflotran} is used to numerically integrate the governing flow equations and obtain pressure values and volumetric flow rates throughout each DFN. 
{\sc dfnTrans} is used to determine pathlines through the DFN and simulate solute transport than can consider matrix diffusion as well as purely advective transport. 
Details of the suite, its abilities,  applications,  and references for detailed implementation are provided in~\citeA{hyman2015dfnworks}.

We consider a semi-generic network loosely based on the fractured and well characterized crystalline rock formation in Forsmark, Sweden, which is a potential host formation for spent civilian nuclear fuel~\cite{follin2014,hartley,joyce2014multiscale,nordqvist2008groundwater}. 
In particular, we select the north-east, north-west, and horizontal family from table  6-75 in~\citeA{skb2010}. 
These three families have the highest fracture intensity values; we choose not to include the north-south and east-west families. 
Generation parameters for the three families are provided in Table~\ref{tbl:dfn_params}.
Our domain is a 200 meter cube. 
Each fracture is represented by a disc with aspect ratio of one, i.e., a circular polygon. 
Fractures are represented as planar discs whose radii $r$ follow a power-law distribution with decay exponent $\gamma$, which is a commonly observed property of fracture networks in the field~\citeA{bonnet2001scaling}.
For computational purposes, we use a truncated power-law distribution with upper and lower  cutoffs ($r_u$; $r_0$) and exponent $\gamma$:
\begin{equation}\label{eq:radius_eq}
p_r(r,r_0,r_u) = \frac{\gamma}{r_0} \frac{(r/r_0)^{-1-\gamma}}{1 - (r_u/r_0)^{-\gamma}}.  
\end{equation}
The fracture centers are uniformly distributed throughout the domain. 
The fracture orientations are sampled from a three dimensional von Mises Fisher distribution,
\begin{equation}\label{eq:fisher}
f({\bf x}; {\boldsymbol \mu}, \kappa ) = \frac{ \kappa \exp( \kappa {\boldsymbol \mu}^{T} {\bf x} )}{ 4 \pi \sinh(\kappa)}~.
\end{equation}
In \eqref{eq:fisher},  ${\boldsymbol \mu}$ is the mean direction vector of the fracture family defined in terms of the families trend and plunge, $T$ denotes transpose, and $\kappa \geq 0$ is the concentration parameter that determines the degree of clustering around the mean direction. 
Values of $\kappa$ close to zero lead in a uniform distribution of points on the sphere while larger values create points with a small deviation from mean direction. 
The distribution is sampled using the method detailed in \cite{wood1994simulation}. 

Variation between the hydraulic properties of the fractures is included using the following relationship between the fracture radius with the transmissivity 
\begin{equation} \label{eq:transmissivity}
    \log(T) = \log(\alpha \cdot r^\beta) + \sigma \cdot Z,
\end{equation}
where $Z \sim \mathcal{N}(0,1)$.  
This relation is referred to as a semi-correlated relationship and is one the models proposed in~\citeA{skb2010}. 
A number of studies have looked into this relationship compared to others, cf.~\citeA{hyman2016fracture} and references therein. 
We use \eqref{eq:transmissivity} as some authors regard it as the most realistic to real-world data, although it comes with the additional computational burden of learning three parameters \cite{follin2014}.
The fracture aperture and permeability are uniform within each fracture and are inferred from the transmissivity via the cubic law in the usual way~\cite{witherspoon1980validity}.

Fractures are placed into the domain from each family until the target fracture intensity $P_{32}$ [m$^{-1}$], values are obtained. 
We adopt the definition of $P_{32}$ proposed in ~\citeA{dershowitz1992interpretation}, where $P_{32}$ is the sum of the surface areas of each fracture ($S_f$) divided by the volume of the domain,
 \begin{equation}\label{eq:p32}
P_{32} = \frac{1}{V}\sum_f  S_f\,.
\end{equation}

For our example, we select five parameters to sample from ranges and keep the remaining generation parameters fixed for all network realizations. 
Our sampled parameters are the decay exponent $\gamma$, fracture intensity $P_{32}$, $\alpha$, $\beta$, and $\sigma$. 
These ranges are based off of known ranges for crystalline rock \cite{brace1980,bonnet2001scaling}. 
Ranges and fixed values are provided in Table~\ref{tbl:dfn_params}.

\begin{table}[t] \label{tbl:dfn_params}
\begin{tabular}{ |p{3cm}||p{3cm}|p{3cm}|p{3cm}|  }
 \hline
 \multicolumn{4}{|c|}{Network Generation Parameters} \\
 \hline
Parameter & Family 1 & Family 2 & Family 3\\
 \hline
 P$_{32}^{\ast}$  & [5 $\cdot 10^{-2}$, 0.2] & [5 $\cdot 10^{-2}$, 0.2]  & [5 $\cdot 10^{-2}$, 0.2] \\
 $\gamma^{\ast}$ [-]  & [ 2.25, 3.5]    &[ 2.25, 3.5]  &   [ 2.25, 3.5]  \\
 $r_0$ [m]  &   10  & 10   & 10\\
 $r_u$ [m]  &   50  & 50   & 50\\
"" $\kappa$    &   14.3  & 12.9  & 15.2 \\
 Trend [$^{\circ}$]   &   329 & 60   & 5\\
 Plunge  [$^{\circ}$]   &   2   & 6     & 86\\
 $\alpha^{\ast}$   &   [1 $\cdot 10^{-10}$, 1$\cdot 10^{-8}$]   & [1 $\cdot 10^{-10}$, 1$\cdot 10^{-8}$]    & [1 $\cdot 10^{-10}$, 1$\cdot 10^{-8}$] \\
 $\beta^{\ast}$   &   [0.1, 1.2]   & [0.1, 1.2]     & [0.1, 1.2]  \\
 $\sigma^{\ast}$  &   [0.5, 1.0]   & [0.5, 1.0]   & [0.5, 1.0]\\
 \hline
\end{tabular}
\caption{Network Generation parameters loosely based on table 6-75 in~\citeA{skb2010}.  Parameters with $\ast$ are samples from the ranges provided. }
\end{table}

\begin{figure}[t] 
\centering
\includegraphics[width=0.65\textwidth]{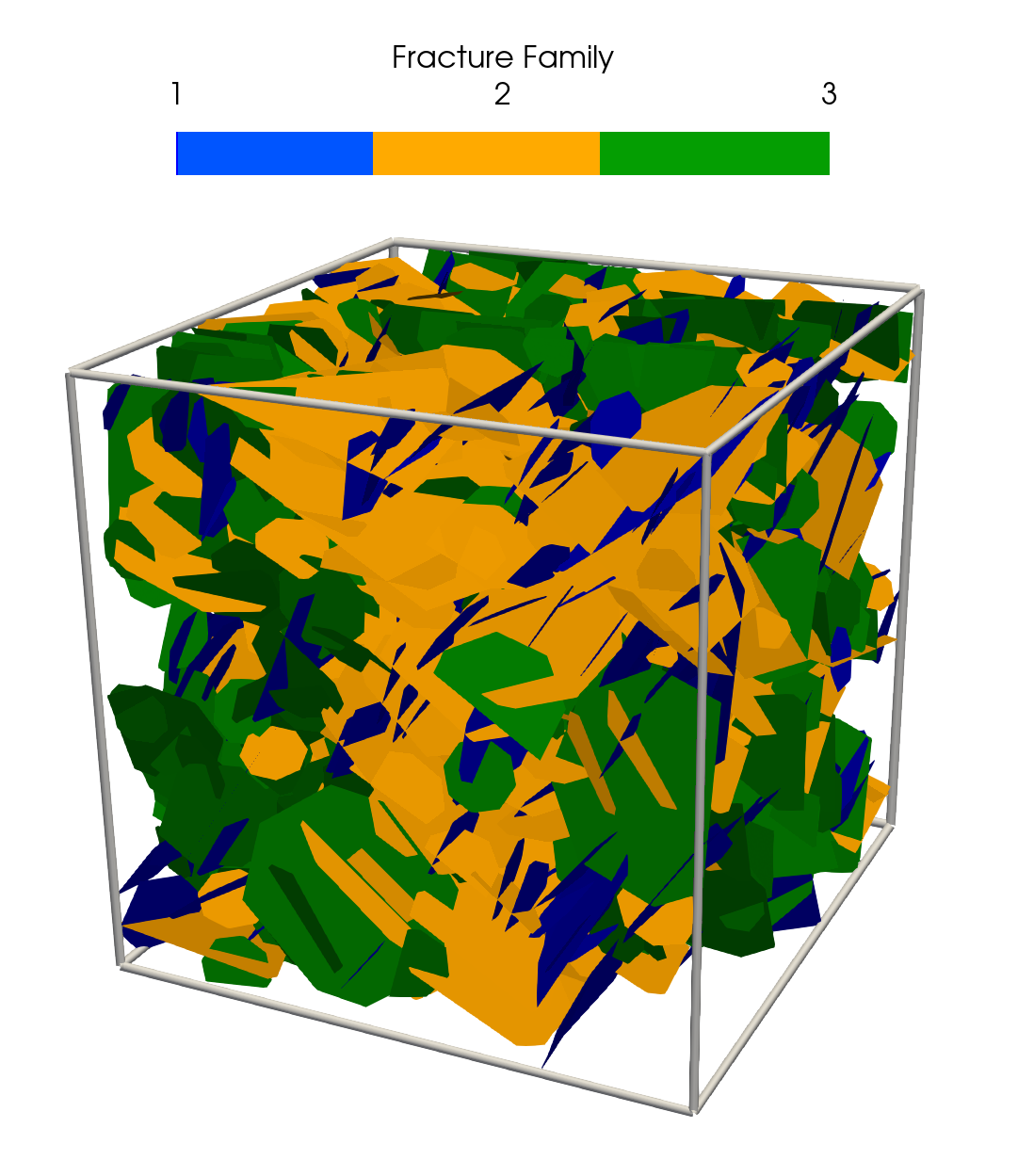}
\caption{One DFN realization. Fractures are colored by family. }\label{fig:dfn}
\end{figure}

\subsection{Flow and Transport Simulations}

We consider the laminar flow of a single incompressible isothermal Newtonian fluid within each fracture network. 
In many subsurface applications, the flow is so slow that inertial effects can be neglected ~\cite{bear1988dynamics,nordqvist2008groundwater,national1996rock,national2021characterization,viswanathan2022from}.
Based on our assumption of aperture uniformity along with the adopted flow properties, flow through the fractures is equivalent to flow between two parallel plates, and thus the Reynolds equations are the equations governing volumetric flow rates and distribution of pressure within the domain~\cite{zimmerman1996hydraulic}.
A pressure difference of 0.1 MPa is prescribed along the x-axis of the networks to drive flow through the domain by assigning fixed pressure values to the entire face of the $+x$ and $-x$ faces of the domain.
No flow boundary conditions are applied along lateral boundaries, and gravity is not included in these simulations.
This model setup creates a single principal flow direction, but the flow within fractures can deviate from this principal direction.

Transport through the network is simulated using a plume of passive, indivisible, and purely advective particles that trace pathlines through the velocity field.
The distribution of particles along the inlet plane is assigned using flux-weighting, which means the number of particles at a position is proportional to the percentage of the total inflow flux occurring at that position~\cite{hyman2015influence,kang2017anomalous,kreft1978on,hyman2021transport}.
One hundred thousand particles are used in each transport simulation. 
Each particle's position $\vx(t;\va)$ evolves according to the advection equation,
\begin{equation}\label{eq:trajectory}
\frac{d \vx(t;\va)}{d t} = \vv[\vx(t)]\,,
\end{equation}
where the Lagrangian velocity $\vv(t)$ is defined in terms of the Eulerian velocity $\vu(\vx)$ as:
\begin{equation}
 \vv(t) = \vu[\vx(t)]\, . 
\end{equation}
When a particle enters the intersection between fractures, wherein the dynamics of flow are a sub-grid scale process and are not fully resolved, we adopt a complete mixing rule where the probability of exiting into a fracture is proportional to the outgoing volumetric flow rate~\cite{Berkowitz1994mass,stockman1997lattice,Park2001intersections,park2003transport,kang2015anomalous,sherman2018characterizing}.
The length of each particle pathline ($\ell$) can be used to parameterize the spatial and temporal coordinates of the particle starting at $\va$ to identify the pathline distance where the particle exits the domain at a position of $x_1 = x_L = 200$ m,
\begin{equation}
\label{eq:lambda}
\lambda(x_L;\va) = \min\{\ell \mid x_1(\ell;\va) \geq x_L\}. 
\end{equation}
In turn, the value of $\lambda(x_L;\va)$ is used to identify the first arrival time $\tau(x_L,\va)$ of a particle at the outlet plane located at $x_L$: 
\begin{equation}
\label{eq:taux}
\tau(x_L;\va) = \int_0^{\lambda(x_L;\va)} \frac{d \ell}{\| \vv_t(\ell) \|}~, 
\end{equation}
 where $\| \vv_t(\ell) \|$ is the magnitude of the Lagrangian velocity along the pathline.
The total mass of the plume $M$, and $\tau(x_L,\va)$ are combined to compute the relative solute mass flux $\psi(t,x_L)$ breakthrough at a time $t$, 
\begin{equation}\label{eq:pdf}
\psi(t,x_L) = \frac{1}{M} \int\limits_{\Omega_a} d \va\delta[\tau(x_L,\va) - t],
\end{equation}
where $\delta(t)$ is the Dirac delta function.
Likewise, we can define the cumulative relative mass flux $\Psi(t,x_L)$ breakthrough at a time $t$, 
\begin{equation}\label{eq:cdf}
\Psi(t,x_L) = \frac{1}{M} \int\limits_{\Omega_a} d \va H[\tau(x_L,\va) - t],
\end{equation}
 where, $H(t)$ is the heavy-side function.
Under the inflow conditions considered here, all particles are assigned the same mass, but their initial positions are distributed according to the inflow flux.

For the purpose of this analysis, we take the $10^{\text{th}}$ percentile of \eqref{eq:cdf} as the scalar output for each data observation.  
Initial experiments have shown that the $10^{\text{th}}$ percentile is a potentially predictable quantity for these sorts of simulations; in contrast, quantities such as the first breakthrough time exhibit such high variance that there is less hope for modeling them explicitly.  
Future extensions of this work may examine other percentiles, or model the entire breakthrough curve as a functional output. 

Initially, five hundred high-fidelity simulations were generated and 487 of them ran to completion (the remaining 13 experienced computational errors during runtime and did not complete).
The input parameters for each of these simulations were chosen according to a Latin Hypercube Sample (LHS) design, which is a pseudo-random sample that explores all areas of the parameter space.  
A LHS is chosen since a strictly random sample may leave areas of the input space unexplored, which would lead to poor emulation for the GP in these areas.  
From these breakthrough data, the $10^{\text{th}}$ percentile of the log breakthrough time is recorded as the response.  There are thus 5 input parameters ($p=5$) to explain a single, scalar response value.  

Following this initial LHS, 250 additional design points were selected according to a sequential design (see Section \ref{sec:seq_design}).  
Every simulation taken at these 250 design points completed without errors.  
As mentioned previously, the sequential design uses diagnostics from emulators fit from a current set of data to determine areas that will most benefit from further simulation runs.  
An additional LHS of 100 high-fidelity runs was attempted for use as a test set, 93 of which completed.  One major outlier was investigated in this set of 93 and ultimately removed, leaving a test set of 92 data values in total.

\section{Uncertainty Quantification Methodology}\label{sec:methodology} In this section, we introduce the three primary statistical tools used to perform a SA for DFN simulations.  First, we propose heteroskedastic joint GPs as a strong model choice to emulate the innate stochasticity observed in DFN simulations.  Next, we introduce a sequential design to strategically choose sample locations to perform additional simulation runs so as to improve our model's fidelity to the observed outputs.  Lastly, we introduce Sobol' Indices as a means to attribute uncertainty to specific input parameters, and discuss ways to quantify and (potentially) explain aleatoric noise.

\subsection{Heteroskedastic Joint Gaussian Processes} \label{subsec:het_joint_gauss_process}

\begin{figure}[t] 
\centering
\includegraphics[width=0.9\textwidth]{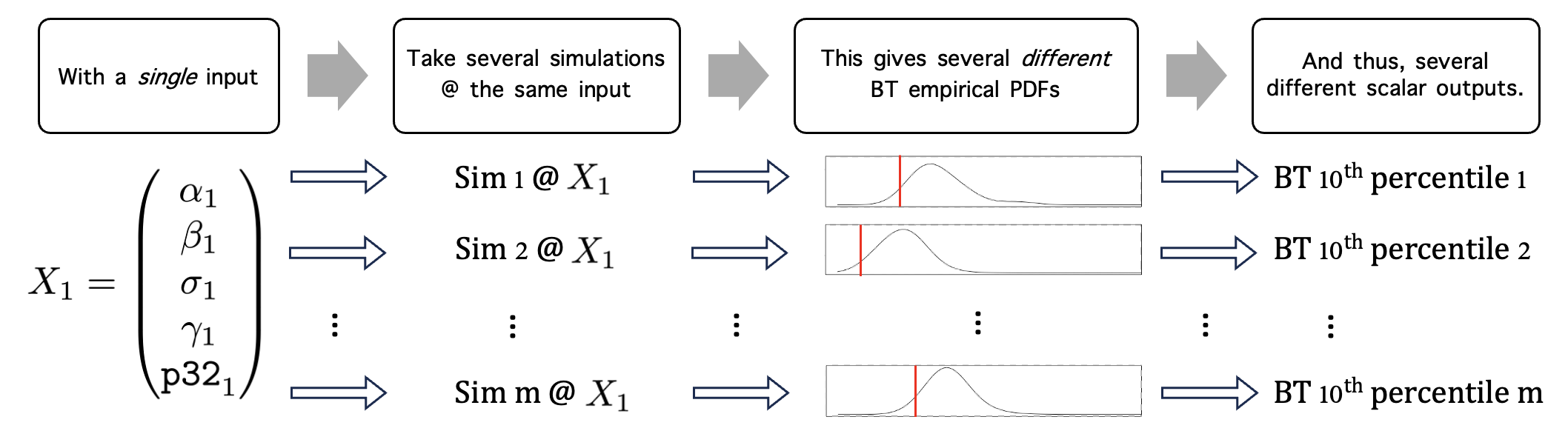}
\caption{Flow chart of how $m$ simulations taken at identical input parameter locations may yield $m$ distinct scalar outputs. }\label{fig:hetero_overview}
\end{figure}


\begin{figure}[t] 
\centering
\includegraphics[width=0.9\textwidth]{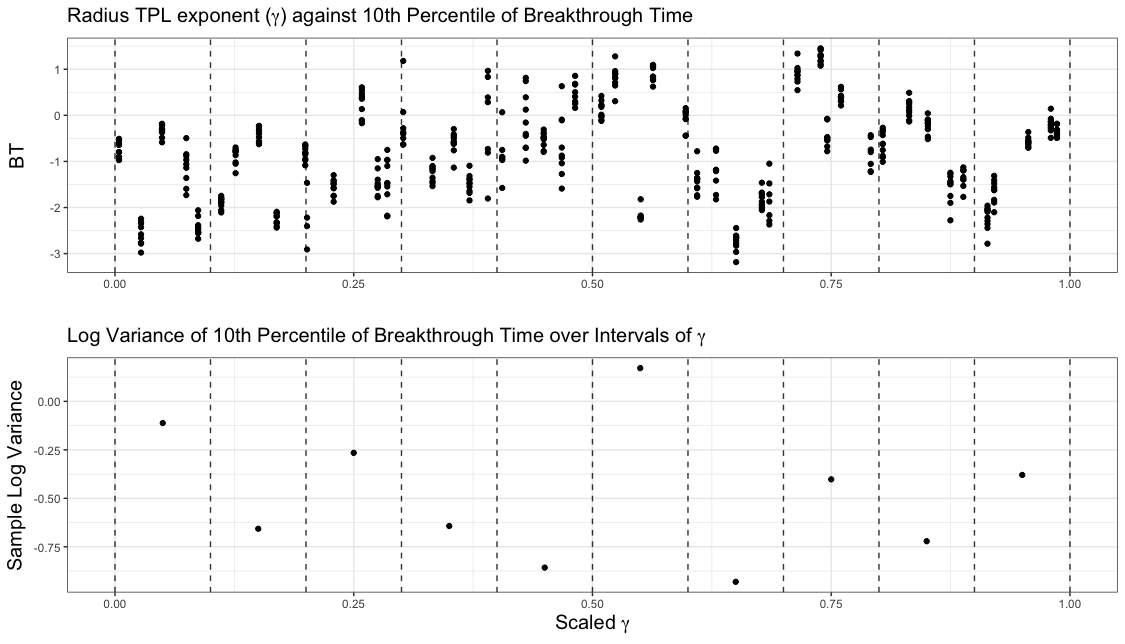}
\caption{Breakthrough times over the scaled $\gamma$ parameter (top) and the log sample variances of these breakthrough times over intervals of 0.1.  The dashed lines mark the intervals from the raw data that are used to calculate the subsample log variances in the bottom graph.  There are significant changes in variance over the input space; while this is evidence of heteroskedasticity, it need not mean that $\gamma$ and the noise are correlated.}\label{fig:hetero_analysis}
\end{figure}

One of the primary modeling challenges in this analysis is how to handle the innate stochasticity observed in DFN simulations.  
In Figure \ref{fig:hetero_overview}, we visualize how several simulations performed at a single input parameter location may lead to several scalar outputs.  This process is also observed in Figure \ref{fig:hetero_analysis}, where the (log) variance in the particle transport data is plotted over several values of the $\gamma$ parameter.  Observe that there are often several points stacked vertically at individual values of $\gamma$; given that the remaining parameters are held fixed at these locations, this shows how identical input values may lead to several different BT values.

Figure \ref{fig:hetero_analysis} further shows how the sample variance changes for different values of $\gamma$, which indicates that the background noise changes over the input space.  This heteroskedasticity is a major challenge to emulating DFN simulations, since many existing emulation methods assume constant noise.  
A method first proposed by \citeA{goldberg1997}, that has since been adopted by several researchers, addresses this issue by modeling the mean and dispersion of the computer code jointly.  That is, two models, $Y_m$ and $Y_d$ are fit, where 
\begin{align}
Y_m(\mathbf{X}) &= \E(Y | \mathbf{X}), \\
Y_d(\mathbf{X}) &= \var(Y | \mathbf{X}) = E( (Y - Y_m(\mathbf{X}))^2 | \mathbf{X}), \label{eq:dispersion_term}
\end{align}
and $Y(\cdot)$ denotes our stochastic simulator that we evaluate at design points $\mathbf{X} = \{ \mathbf{x}_1, \dots, \mathbf{x}_N \}$, $N \in \mathbb{N}$, $\mathbf{x}_i \in \mathbb{R}^p,$ that yield the responses $\mathbf{y}:= \begin{pmatrix} y_1 & \dots & y_N \end{pmatrix}^T, y_i \in \mathbb{R}$.  
For stochastic simulators, this joint model more faithfully emulates the simulator process, since it is capable of quantifying the effects of the input parameter while also learning how the noise changes over the input space.  

Joint heteroskedastic modeling of a stochastic simulation can be performed with various types of models, such as Generalized Linear Models (GLMs) \cite{lee2003}, Generalized Additive Models (GAMs) \cite{iooss2009}, or GPs \cite{marrel2012}.  For this application, we choose to use the joint GP modeling method proposed by \citeA{binois2018}, for which there is the well-regarded R package \texttt{hetGP} \cite{binois2021}.  GPs have several advantages that make them a strong choice for the emulation needs of this analysis: they are nonparametric, the amount of smoothness of the nonlinear functional fit can be controlled via the choice of kernel function, predictive uncertainty is available in closed form because of conditional normal distribution properties, and they can be fit reasonably fast provided sample sizes are relatively small.  In \citeA{marrel2012}, GPs are directly compared against GLMs and GAMs; the authors of this paper concluded that GPs were the only models able to reasonably emulate both $Y_m$ and $Y_d$.  Moving forward, we assume the reader is somewhat familiar with GP modeling (a thorough overview can be found in \citeA{gramacy2020}), so we begin with a review of how to perform joint modeling on the mean and dispersion components of simulation code, borrowing some notation from \citeA{kennedy2023} and \citeA{binois2018}.  We then provide an overview of the updates to this methodology developed by \citeA{binois2018} and implemented as an R package by \citeA{binois2021}.

Let $\lambda^2(\cdot)$ be the noise (or nugget) of $Y(\cdot)$, then
\begin{align*}
Y_m (\cdot) := Y(\cdot) | \lambda^2(\cdot) &\sim \mathcal{N} \left\{ m(\cdot), \sigma^2(\cdot) \right\},
\end{align*}
where the mean and covariance functions are defined according to the kringing equations,
\begin{align}
\text{mean function:} ~~~~~
m(\cdot) =&~ \mathbf{k}(\cdot)^T \mathbf{K}^{-1} \mathbf{y}, \label{eq:mean_func} \\
\text{variance function:} ~~~~~ \sigma^2 (\cdot) =&~ \nu\left( C(\cdot, \cdot) - \mathbf{k}(\cdot)^T \mathbf{K}^{-1} \mathbf{k}(\cdot) \right). \label{eq:var_func}
\end{align}
In the above, $\nu$ is a scaling hyperparameter, and 
\begin{align*}
    \mathbf{k}(\cdot) &= \begin{pmatrix} C(\cdot, \mathbf{x}_1) & \dots & C(\cdot, \mathbf{x}_N) \end{pmatrix}^T, \\
    K_{i,j} &= C(\mathbf{x}_i, \mathbf{x}_j) + \delta_{ij} \lambda^2(\mathbf{x}_i),
\end{align*}
where $K_{i,j}$ is the $(i,j)^{\text{th}}$ element of the matrix $\mathbf{K}$, $\delta_{ij}$ is the Kronecker delta, and $C(\cdot, \cdot)$ is the kernel function of the GP.  While many options exist for the function $C$, we use the Mat\'ern kernel,
\begin{equation} \label{eq:matern_kernel}
C(x_i, x_j) = \prod_{k=1}^p \left(1 + \frac{\sqrt{3}(x_{i,k} - x_{j,k})}{\theta_k}  \right) \exp \left( - \frac{\sqrt{3} (x_{i,k} - x_{j,k})}{\theta_k} \right), 
\end{equation}
where the $\theta$s are the lengthscale hyperparameters, which control the rate of decay of the kernel function.

The Mat\'ern kernel in \eqref{eq:matern_kernel} (which is a Mat\'ern with the smoothness parameter set to $3/2$) enforces less smoothness than the common Gaussian kernel.  For a further discussion of the benefits of this kernel, and why it might be preferable to a Gaussian, see \citeA{stein1999}.  This choice of kernel is called an anisotropic kernel since there are $k$ lengthscale hyperparameters (one for each input) instead of a single, constant lengthscale scalar.  This type of kernel is particularly flexible since it separately controls the lengthscale corresponding to each of the $k$ input features, allowing the GP to learn separate dependence scales for each input.  However, this comes with the burden of learning $k-1$ more hyperparameter values than what is needed for the non-anisotropic case.  Fitting the GP used for this analysis on $Y(\cdot)$ amounts to learning the hyperparameters $H = \{ \nu, \theta_1, \dots, \theta_p \}$ that maximize the GP log marginal likelihood.

The primary difference between the above setup and a homoskedastic GP setup is that the the noise term is assumed to vary across the input space.  For a homoskedastic GP, $\lambda^2$ is either learned via maximum likelihood, or by placing a prior on it (commonly, a inverse gamma prior with constant hyperparameters).  In contrast, for the heteroskedastic case the noise term is given a GP prior:
\begin{equation} \label{eq:noise_emulator}
Y_d (\cdot) := \lambda^2(\cdot) \sim \mathcal{N} \left\{ m_V(\cdot), \sigma^2_V (\cdot) \right\}.
\end{equation}
The mean and covariance functions $m_V$ and $\sigma_V^2$ are defined the same as in \eqref{eq:mean_func} and \eqref{eq:var_func}, but with their own unique set of hyperparameters $H^V$, and the matrix $\mathbf{K}$ replaced by $\mathbf{K}^V$, where
\[ K_{i,j}^V = C_V(\mathbf{x}_i, \mathbf{x}_j) + \delta_{ij} \lambda_V^2, \]
where $\lambda_V^2$ is a constant noise term, which must now be learned like the rest of the hyperparameters, and $C_V$ is defined the same as in \eqref{eq:matern_kernel}.  Fitting the whole joint process now equates to learning $H$, $H^V$, and $\lambda_V^2$.  Learning this many hyperparameters can be challenging for GPs, since they often viewed as ``data hungry'' models, i.e. they scale poorly with the number of data points.  Below we discuss a means, as proposed by \citeA{binois2018}, to help alleviate this issue using replication.

\citeA{binois2018} provide two major updates to the above joint GP methodology.  Their first update involves joint GP modeling in the presence of replication.  Suppose that of the $N$ design points, only $n \ll N$ are unique.  One could greatly save on computational cost, and on the number of hyperparameters that would need to be learned, by working with the lower dimension $n$ and with the averaged responses $\bar{\mathbf{y}} = \begin{pmatrix} \bar{y}_1 & \dots & \bar{y}_n \end{pmatrix}^T$.  Something similar to this is done in \citeA{ankenman2010}: only the mean response GP is fit, and the individual nugget terms are set to $\lambda^{2}(\mathbf{x}_i) := \hat{\lambda}_i^2$, where $\hat{\lambda}_i^2$ is a Method of Moments (MOM) estimator using all repeated observations at $\mathbf{x}_i$.  While elegant in its simplicity, this approach requires that there be sufficient replications so that the MOM estimates are reasonably accurate.  \citeA{binois2018} offer a more satisfying methodology using the Woodbury identity to evaluate the (full-data, naive to replication) equations \eqref{eq:mean_func} and \eqref{eq:var_func} using only the unique $n$ data points and their averaged responses.  Depending on the ratio of $n$ to $N$, this trick can result in dramatic computational savings, since it essentially allows for every ``full-$N$'' calculation to be performed by the ``unique-$n$'' analogue.  Of course, this gain is dependent on the presence of replications, which would mean that sparse computational resources may need to be spent on identical input locations.  This is discussed further in Section \ref{sec:seq_design}.

The second update suggested by \citeA{binois2018} is to fit the second GP not on $\lambda^2 (\cdot)$, by on a smoothed transformation of $\lambda^2 (\cdot)$, which improves interpolation for out-of-sample observations.  In the context of the SA described in Section \ref{subsec:sobol}, this means that any uncertainty attributed to the simulation input parameters from the noise process comes from these transformed values -- not from the raw nugget term $\lambda^2(\cdot)$.  This is elucidated further in \ref{a:smooth_transform}.

We fit the hyperparameters $H, H^V,$ and $\lambda_V^2$ via Maximum Likelihood Estimation (MLE), leveraging several of the MLE design decisions made by \citeA{binois2018} that help facilitate speed and accuracy under the \texttt{hetGP} setup. To ensure that the MLE fit of our GP is reasonable, we look at several validation methods to ensure that the joint model has good coverage and continues to obey the normality assumptions needed for these models.  Once the GP is validated, we perform bootstrapping to quantify the uncertainty on the Sobol' indices (see Section \ref{subsec:sobol}).  This process involves resampling the data, which gives a potentially unique MLE for every bootstrap; thus, the uncertainty on the MLE is simulated and propagated forward to the final uncertainty statements on the Sobol' indices.  This bootstrapping method quantifies the uncertainty on the main statements we wish to make in this analysis.  A Bayesian approach such as Markov Chain Monte Carlo (MCMC), on the other hand, could additionally provide robust uncertainty quantification (UQ) on the hyperparameters themselves. However, this UQ is not the focus of the paper and would come at an exorbitant additional computational burden. For future experiments where we wish to more fully understand the final GP fit, we may consider MCMC as an alternative option, even with its computational difficulties.


\subsection{Replication and sequential design} \label{sec:seq_design}
In a stochastic simulation where identical inputs may lead to different outputs, there is an argument for spending several of a limited pool of simulation runs on the same input locations.  Using simulation runs in this fashion comes at the cost of more thoroughly exploring the input parameter space, but it may provide valuable information on the aleatoric uncertainty.  The statistics research community seems somewhat split on which is more appropriate.  In \citeA{marrel2012}, the approach from \citeA{ankenman2010} is compared against a no-replication joint GP with the aim of estimating the parameters in the Ishigami function \cite{homma1996}.  Using this experiment, the authors conclude that higher accuracy for the mean and dispersion components is achieved without replication.  Meanwhile, \citeA{binois2019} argue that replication in general may have great benefits, and that only the ``fully batched'' version approach used in \citeA{marrel2012} is at fault.  \citeA{binois2019} develop an optimization criterion that chooses new design points sequentially, where these can be either new input locations or replicates at existing input locations.  They conclude that it is appropriate to replicate whenever this choice results in an overall reduction in their criterion, the Integrated Mean Squared Predictive Error (IMPSE).  The IMPSE is the predictive variance for a new value (given all current data points) integrated over the entire input space, which is available in closed-form for GPs \cite{binois2021}.  Thus, minimizing this criterion is a means to decrease the overall predictive variance in the final GP fit.  A second argument for replication is the above update from \citeA{binois2018} on mapping the unique-inputs GP to the full process, since it leads to overall computational savings.

For the purposes of the analysis in this paper, replication has an additional justification.  In the context of a sequential design, where new input locations must be chosen -- and the simulation run -- one at a time, it makes sense to run several simulations at the same location in parallel.  Depending on computational resources, taking batches of simulations at each new input selected by the sequential design would take around the same time as simply performing a single simulation.

We will allow for replication whenever it means a reduction in IMSPE, or whenever it is computationally convenient (due to parallelization).  This leads to a natural ``batching'' sequential design for choosing design points after our initial LHS.  Each new input location is chosen such that it minimizes the IMSPE.  At this location, we run 5 simulations in parallel, add these to the full dataset, and repeat the process.  We repeated this sequential process until (at least) an additional 250 data points were collected. The number of replications at each location was chosen to be 5 simply due to the computational limitations.  In future experiments, we may attempt more replications, especially when we develop the bias-correction approach discussed in Section \ref{sec:discussion}.

\subsection{Sobol Indices for Stochastic Simulations} \label{subsec:sobol}

Sobol indices were originally developed (and are often used) for deterministic models, i.e., models where identical inputs lead to the same output (see, for instance, \citeA{oakley2002}).  The simulation discussed in this paper is stochastic rather than deterministic, meaning that identical inputs may lead to multiple outputs.  \citeA{marrel2012} reduce the stochastic case to the deterministic case by introducing a new variable--called the \textit{seed variable}--to account for the aleatoric uncertainty.  Let $X_\epsilon$ be the seed variable and $\mathbf{X}$ be our vector of observable inputs.  We now wish to study the extended model,
\begin{equation} \label{eq:extended_model}
Y:=f(\mathbf{X}, X_\epsilon) \mapsto \rr.
\end{equation}

The seed variable is neither observable nor controllable, which complicates a formal variance decomposition on \eqref{eq:extended_model}.  It is not immediately clear how one might attribute variation in the output to variation in an input when changes in the seed variable are not observed (and, indeed, changes in an imagined seed variable may not even have contextual meaning).  However, it is still possible to decompose the variance in terms of the general effects of this new variable.

To perform an SA with a seed variable, \citeA{marrel2012} decompose the overall variance of the response $Y$ in terms of the joint GP using the law of total variance,
\begin{align} 
\var(Y) &= \var(\E(Y | \mathbf{X})) + \E( \var (Y | \mathbf{X})) \notag \\
&= \var( Y_m(\mathbf{X}) ) + \E(Y_d(\mathbf{X})) \label{eq:totalvar}.
\end{align}
The term $\E(Y | \mathbf{X})$, which is the mean response given all observable inputs, is understood as the underlying deterministic signal in the stochastic simulation with the seed variable integrated out.  When this mean response is modeled by the GP $Y_m(\mathbf{X})$, we are able to directly analyze the portion of the overall variance of $Y$ that is not caused by changes in the seed variable, and attribute it to the different observable inputs.

An analysis of $Y_m(\mathbf{X})$ is undertaken using the Sobol' decomposition \cite{sobol2001}.  This method decomposes $Y_m$ according to its interaction effects so that,
\[ Y_m(\mathbf{X})  := g_0(\mathbf{X}) + \sum_{i=1}^p g_i (\mathbf{X}) + \sum_{i=1}^p \sum_{j > i} g_{ij} (\mathbf{X}) + \dots + g_{1 \dots p} (\mathbf{X}). \]

Each of the above terms is constructed so that it is centered at zero and orthogonal to all other terms:
\begin{align*}
    g_0 &= \int Y_m(\mathbf{X}) p(\mathbf{X}) d\mathbf{X} \\
    g_i (x_i) &= \int Y_m(\mathbf{X}) p(\mathbf{X}) d\mathbf{X}_{-i} - g_0 \\
    g_{ij} (x_i, x_j) &= \int Y_m(\mathbf{X}) p(\mathbf{X}) d\mathbf{X}_{-ij} - g_0 - g_i(x_i) - g_j(x_j),
\end{align*}
and so on, assuming that $\mathbf{X}$ is independent and uniformly distributed.  The properties of this construction (zero-centered and orthogonal) imply that the variance of $Y_m$ can be similarly decomposed linearly,
\begin{equation} \label{eq:var_decomp_sobol}
    \var(Y_m) = \sum_{i=1}^p V_i(Y_m) + \sum_{i=1}^p \sum_{j>i} V_{ij}(Y_m) + \dots + V_{1\dots p}(Y_m),
\end{equation}
where $V_J(Y_m) = \var(g_J(\mathbf{X})),$ for some index set $J \subset \{1,\dots,p\}$.  

The Sobol' index is defined as the proportion of total variance that can be attributed to each element in \eqref{eq:var_decomp_sobol},
\[ S_J = \frac{V_J(Y_m) }{\var(Y_m)}. \]
The case of $J=i$ is called the \textit{First-Order effect} of $i$, while all other cases are called the \textit{Interaction Effects}.  The \textit{Total Effect} of $i$ is the sum,
\[ S_{T_i} = \sum_{J \subseteq \{1,\dots,p\} : i \in J} S_J \]

Sobol' indices are the tools used in this paper to perform an SA on the mean response of the DFN ensemble, since they allow one to attribute observed uncertainty to the individual simulation parameters and their interactions.  Since these indices do not admit closed form solutions for their corresponding integrals, we calculate them using Monte Carlo sampling, following the recommendations found in \citeA{saltelli2010}.  Note that these approximations are calculated using $Y_m$ as an emulator for the mean response $\E(Y|\mathbf{X})$; this choice was made because it is unclear how to simulate values from $\E(Y|\mathbf{X})$ directly and because full runs of the DFN model are of a computational cost that would make Monte Carlo sampling of this quantity infeasible.

An SA on the remaining variance $\E(\var(Y | \mathbf{X}))$ is a much more elusive prospect for several reasons.  Although the above process admits a GP fit to the dispersion term, an SA on $Y_d(\mathbf{X})$ would be equivalent to a decomposition of $\var(Y_d(\mathbf{X})),$ \textit{not} on the $\E(Y_d(\mathbf{X}))$ term of interest.  Furthermore, as mentioned earlier, such an analysis is innately one on changes in the seed variable, which is an imagined concept without any meaningful notion of distance.

While a classical SA on the aleatoric uncertainty is theoretically cumbersome with the presence of a seed variable, we nonetheless remain interested in how changes in our observable inputs lead to changes in the underlying noise of DFN simulations.  Fortunately, there are several deductions on this stochasticity one can make using a joint GP.  Consider the variance decomposition from \eqref{eq:var_decomp_sobol} using the full model \eqref{eq:extended_model} that includes the seed variable: 
\begin{align}\label{eq:var_w_seed}
\var(Y) = V_\epsilon(Y) + \sum_{i=1}^p \sum_{|J|=i} \left[ V_J(Y) + V_{J\epsilon}(Y) \right].
\end{align}
Using the law of total expectation, the terms not associated with the seed variable are equivalent to the first term in \eqref{eq:totalvar},
\[ \var( Y_m(\mathbf{X}) ) =  \sum_{i=1}^p \sum_{|J|=i} V_i(Y_m) =  \sum_{i=1}^p \sum_{|J|=i} V_i(Y). \]
Thus, combining the above with \eqref{eq:totalvar} and \eqref{eq:var_w_seed},  
\begin{equation} \label{eq:seed_variance}
\E( Y_d(\mathbf{X}) ) =  V_\epsilon(Y) + \sum_{i=1}^p \sum_{|J|=i} V_{J\epsilon}(Y).
\end{equation}
From here, the usual calculation for the Total Effect of the seed variable is derived simply by dividing by the total variance:
\begin{equation} \label{eq:seed_total_effect}
S_{T_\epsilon} = \frac{V_\epsilon(Y) + \sum_{i=1}^p \sum_{|J|=i} V_{J\epsilon}(Y)}{\var(Y)} = \frac{ \E( Y_d(\mathbf{X}) ) }{ \var(Y)}. 
\end{equation}
Thus, while an explicit SA on $Y_d(\mathbf{X})$ will not decompose the variance attributed to the system's aleatory uncertainty, we have deduced that whatever proportion of variance not accounted for in an SA on the mean response must make up the Total Effect of the seed variable.

Calculating the First-Order effect of $X_\epsilon$, and the interaction effects between $X_\epsilon$ and the other input variables, are outside the reach of current research \cite{baker2022}.  However, relative statements about these quantities can be made according to an SA on $Y_d$.  While this may seem contradictory to previous statements discouraging an SA on $Y_d$, consider the following: after performing an SA on $Y_d$, we discover that $X_1$ accounts for a quarter of the variation in $Y_d(\mathbf{X})$, $X_2$ accounts for half, and $X_3$ accounts for none of the variation.  While the variance attributed to $X_1$ and $X_2$ here is not from the overall variance in the response (for the reasons discussed above), it nonetheless implies that they each have an effect on $\E(Y_d(\mathbf{X})),$ with $X_1$ having the greater effect.  Thus, we conclude that $S_{1 \epsilon} > S_{2 \epsilon}$.  On the other hand, the lack of variance attributed to $X_3$ implies that $X_3$ has no effect on $\E(Y_d(\mathbf{X})),$ so we conclude that $S_{3 \epsilon}=0$.  Further justification for this sort of analysis can be found in \citeA{iooss2009}, and more thorough examples of this sort of relative analysis can be found in \citeA{marrel2012}.  We will perform a relative analysis of the interaction terms with the seed variable when we discuss the dispersion GP fit to our ensemble.

It is important to note that the presence of heteroskedasticity need not mean that the simulation input parameters are directly correlated with changes in the noise term.  That is, it is possible that changes in the underlying noise of the system (as observed in Figure \ref{fig:hetero_overview}) is itself independent to changes in the inputs.  Above, we have equated this to mean that the interaction Sobol' index between that input parameter and the seed variable is zero.

\section{Modeling Results} \label{sec:experiment}

\subsection{Emulator Validation} \label{subsec:validation}
\begin{figure} 
\centering
\includegraphics[width=0.6\textwidth]{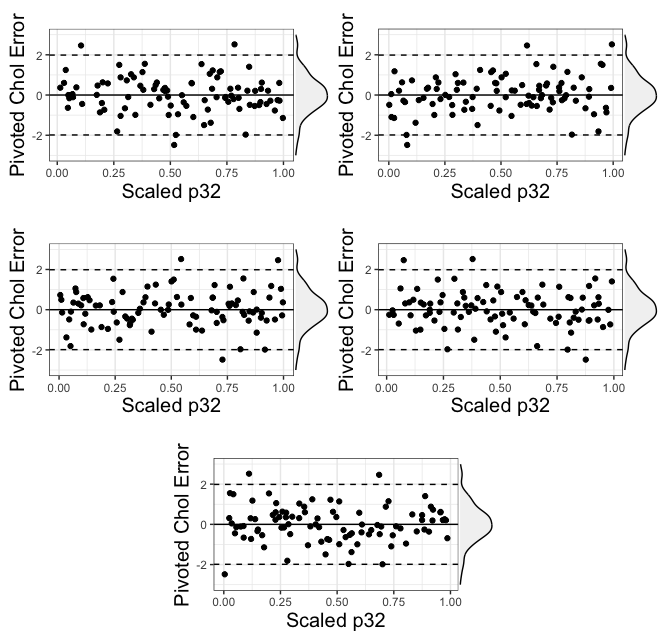}
\caption{Scaled Choleskey Errors on the validation data.  The dashed lines mark the approximately normal theoretical 95\% quantiles and a solid line is place at zero for reference.  Empirical densities of the marginals on the Pivoted Choleskey Errors are drawn on the far right of each plot.} \label{fig:chol_errors}
\end{figure}

Prior to performing the SA, we first assess the fidelity to which the final GP emulates the DFN simulation.  We will take three different approaches to assessing a good model fit for our GPs: graphically examining the pivoted Cholesky errors for approximate normality, as suggested by \citeA{bastos2009}, calculating the Root Mean Squared Error (RMSE) for predictions on the test set, and examining the converage of the predicted confidence intervals on the test set.  The first approach transforms the predictions on the test set using the predicted variance,
\begin{equation} \label{eq:pivoted_chol}
G^{-1} \left( y^* - \E\left( Y_m(\mathbf{X}^*) | \mathbf{y} \right) \right),
\end{equation}
where $(\mathbf{y}^*, \mathbf{X}^*)$ are the data from the test set, and $G$ is the pivoted Cholesky decomposition on the variance of the joint emulator,
\[ \var(Y_m(\mathbf{X}^*) | \mathbf{y} ) + E( Y_d (\mathbf{X}^*)| \mathbf{y} ) = G G^T. \]
Since the emulator used is a GP, the transformed errors in \eqref{eq:pivoted_chol} should follow a Student's $t$-distribution (which will approximately normal for the test set considered here with more than 30 observations), which we will assess by plotting the transformed errors against the theoretical 95\% quantiles.  These quantiles are plotted as red lines in Figure \ref{fig:chol_errors}, while the errors are plotted along the marginals of each of the five variables in this study.  We do not observe any significant patterns in these errors, and we find the approximate normality assumption reasonable.

The final RMSE at the 50th iteration of the sequential design was approximately 0.732.  The RMSE over the course of this design did not see significant changes; it mostly varied above and below this value.  This may be in part due to the error initially being very low.  Critical assessment of this design has been discussed thoroughly in the literature (see Section \ref{sec:seq_design}), but as far as the authors are aware this is the first time it has been applied to DFN simulations.  Further assessment on the merits of this sequential design for this specific experiment and exploration of alternative designs are left as an interesting direction for future research.

\begin{figure} 
\centering
\includegraphics[width=0.6\textwidth]{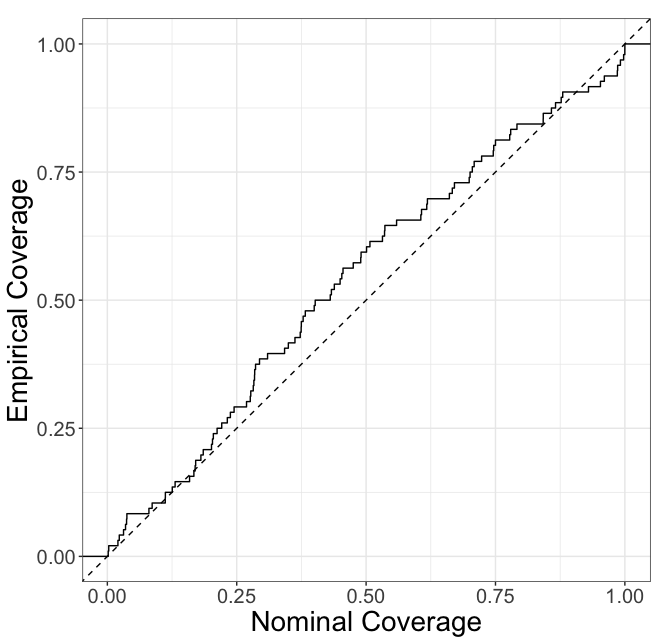}
\caption{Nominal vs. Empirical coverage for the final heteroskedastic GP fit.  This plot shows a strong match between coverages set by a researcher and those observed on the testing set.} \label{fig:predictive_coverage}
\end{figure}

As a final validation criterion, we examine the empirical coverage of the predictive confidence intervals (CI) at every level on the test set.  In Figure \ref{fig:predictive_coverage}, we graphically compare the nominal coverage (the coverage set by the researcher, i.e. a 95\% CI, a 90\% CI, etc.) against the empirical coverage (the coverage we observe).  A perfect fit would have a diagonal line, which would correspond to the case where every possible coverage value set by the researcher led to an identical observed coverage for the testing values.  For instance, a predictive CI with a desired (nominal) coverage of 95\% would align with an observed (empirical) coverage of 95\%, which corresponds to the point on the diagonal dashed line in Figure \ref{fig:predictive_coverage}.  The results in Figure \ref{fig:predictive_coverage} show that the nominal and empirical coverages approximately match, with some minor overperformance for the central coverage values; this further validates the strong fit of the heteroskedastic GP.

\subsection{Sensitivity Analysis on the Joint Emulator} \label{subsec:sa}
We perform two SAs, one explicitly on the mean response GP, $Y_m$, and a second, relative SA on the dispersion GP, $Y_d$, according to the process outlined in \citeA{marrel2012}.  Due to the relatively small size of the input parameter dimension, the Monte Carlo algorithm used to calculate an estimate on the Sobol' indices is quick enough that it is feasible to perform bootstrapping on these values, which will give UQ for these estimates.  The overall setup to calculate confidence intervals on the Sobol' indices is: resample the high-fidelity simulation data (with replacement), fit a \texttt{hetGP} on this new data, then perform a Monte Carlo approximation on the Sobol' indices (as averages of 10000 Monte Carlo samples).  This process was repeated until 5000 bootstrapped averages are calculated.

\begin{figure} 
\centering
\includegraphics[width=0.9\textwidth]{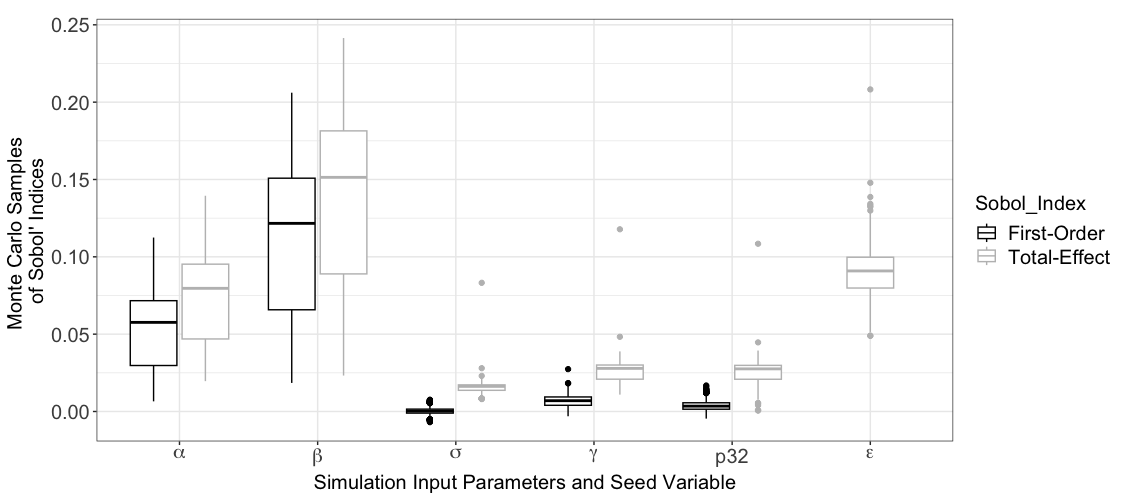}
\caption{Monte Carlo Samples of the First-Order (top) and Total-Effect (bottom) Sobol' indices.  The bootstrapped distributions on these two metrics do not vary significantly between the two plots; this implies that there is little variance attributed to the higher-order interaction terms.  Note that for the total-effect estimations other than the explicit estimate on the Total-Effect for the seed variable $\epsilon$, the seed variable is assumed to be integrated out. } \label{fig:sa_on_mean}
\end{figure}

In Figure \ref{fig:sa_on_mean}, we graph the First-Order and Total-Effect Sobol' indices on our five parameters using only the mean response GP, and the Total-Effect Sobol' index on the seed variable using the entire joint process.  We did not examine higher-order interaction terms because the distributions across these two plots are near-identical.  The parameters $\alpha$ and $\beta$ from \eqref{eq:transmissivity} accounted for a majority of the variance observed in the model, implying that these variables, which govern the relationship between radius size and log-transmissivity, are crucial to explaining the variation in log breakthrough time for the experiment in this paper.  In comparison, the variables that govern the uncertainty in this relationship, the size of the randomly-placed radii, and the density of fractures, accounted for significantly less variation.

The bootstrapped estimates on the Total-Effect Sobol' index for the seed variable $\epsilon$ were calculated using \eqref{eq:seed_total_effect}.  As seen in Figure \ref{fig:sa_on_mean}, there is evidence that that Total-Effect of the seed variable $\epsilon$ is non-zero.  This coincides with the modeling assumptions made in this paper as well as the observations made with Figure \ref{fig:hetero_analysis}: the data have heteroskedastic noise across the input space (since $S_{T_\epsilon}=0$ would indicate homoskedasticity).  We wish to determine if there is evidence for any significant interactions between the seed variable and the other variables, as this would indicate that the value of certain variables may have an effect on the level of aleatoric noise.  To do this, we calculate bootstrapped 95\% confidence intervals on the First-Order Sobol' indices derived using the dispersion GP, $Y_d$.  When zero is in the confidence intervals for variable $i \in \{\alpha, \beta, \sigma, \gamma, \text{p32}\}$, we will say that there is not sufficient evidence that variable $i$ is influential to $Y_d$, and instead accept the conclusion that $S_{i\epsilon} = 0$.  If zero is not in confidence interval $i$, we will then make relative statements about $S_{i\epsilon}$ using the logic described in \citeA{iooss2009}.

The experiment described in the previous paragraph found that zero is present in every confidence interval calculated, so there is insufficient evidence to conclude any interactions between the simulation input parameters and the seed variable.  While there is ample evidence that the aleatory noise varies over the input space, there is not sufficient evidence that changes in any specific variable accounts for any of this noise jointly with the seed variable.  A table of the values in Figure \ref{fig:sa_on_mean} is available in  \ref{a:table_6_values}.

\section{Discussion}\label{sec:discussion}
This paper is a natural application of modern advances in the Statistics and Machine Learning literature.  We have shown that heteroskedastic emulators coupled with a formal SA using Sobol' indices are valuable tools for understanding simulations that have non-constant aleatoric noise.  As far as the authors are aware, this is the first time that a SA for DFN simulators directly addresses heteroskedastic noise.  There have been a recent SA in the literature on DFN simulations \cite{brooks2022sensitivity}, but this paper focuses on peak nuclear waste concentrations in aquifers as a response over time and takes an entirely different approach to emulation that does not account for heteroskedasticity.

The rigorous workflow for characterizing DFN ensemble uncertainty has broad implications for better understanding of a growing body of earth science simulations. Our results provide guidance for focusing data collection on transmissivity parameters, allowing intelligent allocation of resources in real world systems that often are data limited.  Additionally, measurable uncertainty in DFN networks can provide confidence in predictions from gas transport simulations, allowing decision makers to more easily arrive at answers to subsurface gas transport problems ranging from geothermal energy to underground nuclear tests. 

The observation on the importance of the transmissivity parameters also has exciting ramifications for ensuing applied statistical tasks of interest.  For future experiments that try to calibrate \textsc{dfnWorks} to field data, it may make most sense to focus on reducing the input space to just $\alpha$ and $\beta$, since these have the greatest effect on the variation in the output that is not attributed to white noise.  The results in this paper also suggest that while DFNs are highly stochastic simulations, there is nonetheless some signal to be gleaned from changing input parameters that ``overrides'' the noise.

There are several directions for future research.  The scope of this paper was focused by only considering input parameter ranges corresponding to crystalline rock and hydrogen gas.  Future iterations of this work could look at different rock types and different types of gases.  A more significant extension would be to include categorical variables in this analysis, perhaps as a way to see if the type of rock/gas significantly affects the variation in the output.  A similar approach may be able to analyze the number of fracture families added as a discrete variable.

An additional direction for further research would be to incorporate low-fidelity approximations into this analysis, such as the fast 2D graph approximations discussed in \citeA{strait2023}.  With replicates, it would be possible to build a bias correction model at individual points in the input parameter space.  This model would then be able to produce several quick approximations of the high-fidelity data using runs of the low-fidelity approximation, exploiting the use of replicates in heteroskedastic GP modeling.

As a final direction for future research, one could look at the BT curve output as a full functions, rather than just reducing it to its $10^{\text{th}}$ percentile.  This would allow us to explore how these input parameters may influence early vs. late breakthrough in DFNs.

\section*{Open Research Section}
The simulation experiment performed in this paper uses the \textsc{dfnWorks} software, which is publicly available at \url{https://dfnworks.lanl.gov/}.  Specific values to perform the simulations described in this paper are found in Table \ref{tbl:dfn_params}.  The \textsc{R} code to fit the emulators described in this paper is under review at Los Alamos National Lab and should be available in 2024.

\acknowledgments
Research presented in this article was supported by the Laboratory Directed Research and Development program of Los Alamos National Laboratory under project number 20220019DR. Los Alamos National Laboratory is operated by Triad National Security, LLC, for the National Nuclear Security Administration of U.S. Department of Energy (Contract No. 89233218CNA000001).

\section*{Conflict of interest}

The authors have no non-financial or other financial competing interests to declare that are relevant to the content of this article other than the aforementioned declared funding source.

\bibliography{bibliography}       

\begin{appendix}
\section{Gaussian Process Emulation of a Heteroskedastic Noise Term using Latent Variables}\label{a:smooth_transform}

Per the recommendations discussed in Section \ref{subsec:het_joint_gauss_process}, the GP emulator is placed on a smooth transformation of the noise term in \eqref{eq:noise_emulator}.  An SA performed directly on this emulator may not be appropriate for the aim of commenting on the dispersion term in \eqref{eq:dispersion_term}.  In this appendix, we will show that a direct transformation of the samples will allow for a proper SA.

In the following, we will assume that there is some degree of replication.  Let $\mathfrak{X} = \{\mathfrak{x}_1, \dots, \mathfrak{x}_n\}$ be the unique design points, $\{a_1, \dots a_n \}$ be the number of observations at each of these points, and $A_n := \text{diag}(a_1, \dots, a_n)$ be a matrix with these counts along the diagonal.  For the application in this paper, the latent variables $\delta_1, \dots, \delta_n$ are emulated via a GP, where
\[ \Lambda = \mathbf{C}(\mathbf{C} + gA_n^{-1})^{-1}( \Delta - \beta_0 I_n). \]
Here, $\Lambda := \text{diag}(\lambda_1^2, \dots, \lambda_n^2)$, $\Delta := \text{diag}(\delta_1, \dots, \delta_n)$, $\beta_0$ is the mean hyperparameter on the GP prior of $\Delta$, and the elements of the matrix $\mathbf{C}$ are defined as $C_{ij} := C(\mathfrak{x}_i, \mathfrak{x}_j)$.  Let $Y_d^*$ be a GP fit to emulate $\Delta$.  Then according to the above relation,
\begin{align*}
  \var(\mathbf{C}(\mathbf{C} + gA_n^{-1})^{-1}Y_d^*) =  \var(Y_d).
\end{align*}
Since the estimation of the Sobol' indices is done via Monte Carlo sampling, it is reasonable to take samples from $Y_d^*$, feed these through the linear map $\mathbf{C}(\mathbf{C} + g A_n^{-1})^{-1}$, then perform the variance estimates used in the Sobol' indices computations.

 \newpage 
\section{Specific Numbers from Figure \ref{fig:sa_on_mean}} \label{a:table_6_values}

\begin{table}[!htbp]
\begin{center}
\begin{tabular}{||c c c c c||} 
 \hline
  & $S_i$ mean & $S_i$ st. dev. & $S_{T_i}$ mean & $S_{T_i}$ st. dev. \\ [0.5ex] 
 \hline
 $i=\alpha$ & 0.0576 & 0.0231 & 0.0796 & 0.0270 \\ 
 \hline
 $i=\beta$ & 0.1216 & 0.0465 & 0.1514 & 0.0517 \\ 
 \hline
 $i=\sigma$ & 0.0002 & 0.0020 & 0.0162 & 0.0030 \\ 
 \hline
 $i=\gamma$ & .0070 & 0.0037 & 0.0279 & 0.0064 \\ 
 \hline
 $i=$p32 & 0.0034 & 0.0030 & 0.0279 & 0.0064 \\ 
 \hline
 $i=\epsilon$ & NA & NA & 0.0897 & 0.0142 \\ 
 \hline
\end{tabular}
\caption{Sobol' indices estimated using the joint heterskedastic GP.  These numbers correspond to those observed in Figure \ref{fig:sa_on_mean}} \label{tab:sa_on_mean}
\end{center}
\end{table}


\end{appendix}

%
%

%
%


%
%
%
%
%

\end{document}